\documentclass[a4paper,12pt]{article}

\usepackage[dvips]{graphicx}
\usepackage{epsfig}

\begin{document}

\author{C. Barrab\`es\thanks{E-mail : barrabes@celfi.phys.univ-tours.fr}\\
\small Laboratoire de Math\'ematiques et Physique Th\'eorique,\\
\small CNRS/UMR 6083, Universit\'e F. Rabelais, 37200 TOURS,
France\\V. P. Frolov\thanks{E-mail :
frolov@phys.ualberta.ca}\\\small Theoretical Physics Institute,
Department of Physics,\\\small University of Alberta, Edmonton,
Canada T6G 2J1
\\\small and \\P. A. Hogan\thanks{E-mail : peter.hogan@ucd.ie}\\
\small Mathematical Physics Department,\\ \small  National
University of Ireland Dublin, Belfield, Dublin 4, Ireland}

\title{Wave and Particle Scattering Properties of High Speed Black Holes}
\date{}
\maketitle

\begin{abstract}
The light--like limit of the Kerr gravitational field relative to
a distant observer moving rectilinearly in an arbitrary direction
is an impulsive plane gravitational wave with a singular point on
its wave front. By colliding particles with this wave we show that
they have the same focussing properties as high speed particles
scattered by the original black hole. By colliding photons with
the gravitational wave we show that there is a circular disk,
centered on the singular point on the wave front, having the
property that photons colliding with the wave within this disk are
reflected back and travel with the wave. This result is
approximate in the sense that there are observers who can see a
dim (as opposed to opaque) circular disk on their sky. By
colliding plane electromagnetic waves with the gravitational wave
we show that the reflected electromagnetic waves are the high
frequency waves.

\end{abstract}
\thispagestyle{empty}
\newpage

\section{Introduction}\indent
When the Kerr black hole, located at the spatial origin of
coordinates $\{\bar x, \bar y, \bar z, \bar t\}$, has its axis
pointing in an arbitrary direction relative to this asymptotically
rectangular Cartesian coordinate system the metric tensor is given
in Kerr--Schild form by
\begin{equation}\label{1.1"}
\bar g_{\mu\nu}=\bar\eta _{\mu\nu}+2\,H\,\bar k_{\mu}\,\bar
k_{\nu}\ ,\end{equation} with $\bar\eta _{\mu\nu}={\rm diag}(1, 1,
1, -1)$, \begin{equation}\label{1.2"} H=\frac{m\,\bar r^3}{\bar
r^4+({\bf a}\cdot\bar {\bf x})^2}\ ,\end{equation} and
\begin{equation}\label{1.3"}
\bar k_{\mu}\,d\bar x^{\mu}=\frac{({\bf a}\cdot \bar {\bf
x})\,({\bf a}\cdot d\bar {\bf x})}{\bar r\,(\bar r^2+|{\bf
a}|^2)}+\frac{\bar r\,(\bar {\bf x}\cdot \bar {\bf dx})}{\bar
r^2+|{\bf a}|^2}-\frac{{\bf a}\cdot ({\bf \bar x}\times d{\bf \bar
x})}{(\bar r^2+|{\bf a}|^2)}-d\bar t\ .\end{equation}Here ${\bf
a}=(c, b, a)$, with $|{\bf a}|^2=a^2+b^2+c^2$ and $\bar {\bf
x}=(\bar x, \bar y, \bar z)$ with $\bar r$ given in terms of $\bar
x, \bar y, \bar z$ by
\begin{equation}\label{1.4"}
\bar x^2+\bar y^2+\bar z^2+\frac{(c\bar x+b\bar y+a\bar z)^2}{\bar
r^2}=\bar r^2+a^2+b^2+c^2\ .\end{equation} The dot and cross above
signify the usual scalar and vector product in three--dimensional
Euclidean space. This form of the Kerr metric has been given by
Weinberg \cite{Wei} and a derivation is given in \cite{PR04}. When
we consider the deflection of a highly relativistic particle in
the Kerr gravitational field \cite{PR04} we proceed as follows:
The direction of the incoming particle is parallel to the positive
$\bar x$--axis. The particle is projected with 3--velocity $v$ at
a large distance from the black hole and with a non--zero (large)
impact parameter. Since the black hole axis points in an arbitrary
direction, projecting a particle in the $\bar x$--direction is
equivalent to projecting in an arbitrary direction. We call $\bar
S$, with coordinates $\{\bar x, \bar y, \bar z, \bar t\}$, the
rest frame of the Kerr source and $S$, with coordinates $\{x, y,
z, t\}$, the rest--frame of the high speed particle. These frames
are related by the Lorentz boost in the $\bar x$--direction
\begin{equation}\label{1.5"}
\bar x=\gamma\,(x+v\,t)\ ,\qquad \bar y=y\ ,\qquad \bar z=z
,\qquad \bar t=\gamma\,(t+v\,x)\ ,\end{equation}with $\gamma
=(1-v^2)^{-1/2}$. In $S$ the particle sees the Kerr source moving
towards it with a speed close to the speed of light (unity in our
units). In the ultrarelativistic limit the Kerr gravitational
field looks to the particle as the gravitational field of a
plane--fronted impulsive gravitational wave. The line--element of
the space--time model of this field is derived in \cite{PR04} and
can be written in the form
\begin{equation}\label{1.1}
ds^2=dx^2+dy^2+dz^2-dt^2-2\,H\,(dx+dt)^2\ ,\end{equation} with
\begin{equation}\label{1.2}
H=h(y, z)\,\delta (x+t)\ ,\qquad h(y,
z)=2p\,\log\{(y-a)^2+(z+b)^2\}\ ,
\end{equation}with $p, a, b$ constants. The
gravitational field described by (\ref{1.1}) and (\ref{1.2})
results in the limit $v\mapsto 1$ with the mass $m$ of the black
hole tending to zero in this limit in such a way that
$m\,(1-v^2)^{-1/2}\mapsto p$. Here $b, a$ are respectively the
components of the angular momentum per unit mass of the black hole
in the $y$ and $z$ directions. The component of the angular
momentum per unit mass in the $x$ direction drops out in the
light--like limit on account of the way in which it must scale in
terms of $v$ due to Lorentz contraction (this is explained in
terms of the behavior of the multipole moments of the isolated
source in \cite{PR01}, \cite{PR03}). The hyperplane $x+t=0$ is
null and the Riemann tensor of the space--time has components
which are multiples of the Dirac delta function $\delta (x+t)$.
This tensor is Type N in the Petrov classification and $x+t=0$ is
the history of a plane impulsive gravitational wave. As well as
being singular on the null hyperplane $x+t=0$ the Riemann tensor
is also singular at $y=a,\ z=-b$ which is the singularity of the
logarithm in (1.2). This is a null geodesic generator of the
hyperplane $x+t=0$ and is a singular point on the plane wave
front. When $a=b=0$ (\ref{1.1}), (\ref{1.2}) reduces to the
gravitational field of the Schwarzschild source boosted to the
speed of light originally calculated by Aichelburg and Sexl
\cite{AS}.

In this paper we demonstrate in section 2 that \emph{the focussing
behavior} of a time--like congruence in the space--time with
line--element (\ref{1.1}) is exactly what one would expect in the
field of a rotating source. The focussing of geodesics in the
space--time of general plane gravitational waves has been
discussed in \cite{GV}. In section 3 we calculate the behavior of
photons (a null geodesic congruence) which experience a head--on
collision with the impulsive gravitational wave described by
(\ref{1.1}) and (\ref{1.2}) as a prelude to the study of the
collision of plane electromagnetic waves with the gravitational
wave. This enables us to demonstrate in section 4 \emph{the
dimming of the light signal} by the gravitational wave and to
indicate that it is the high electromagnetic wave frequencies
being reflected by the gravitational wave that is responsible for
the dimming. When studying photons and electromagnetic waves in
the present paper we solve the null geodesic equations and
Maxwell's equations in the space--time given by (\ref{1.1}) and
(\ref{1.2}). This is required in order to describe the new signal
dimming phenomenon which applies to a non--rotating as well as to
a rotating black hole.

\setcounter{equation}{0}
\section{Scattering of Particles:Time-Like Geodesics}\indent
The time--like geodesic equations in the space--time with
line--element (\ref{1.1}) with (\ref{1.2}) can be used to
calculate the angles of deflection of a high speed particle
projected in any direction into the field of the black hole \cite
{PR04} \cite{CQG}. This involves two approximations: in the rest
frame of the incoming particle the Kerr field is taken to be given
approximately by (\ref{1.1}) with (\ref{1.2}) and the impact
parameter must be large. The leading terms in the calculated
deflection angles are found to coincide with the angles of
deflection of photons projected into the Kerr field \cite{PR04}.
\begin{figure}[h!]
\begin{center}
\includegraphics[width = 12cm]{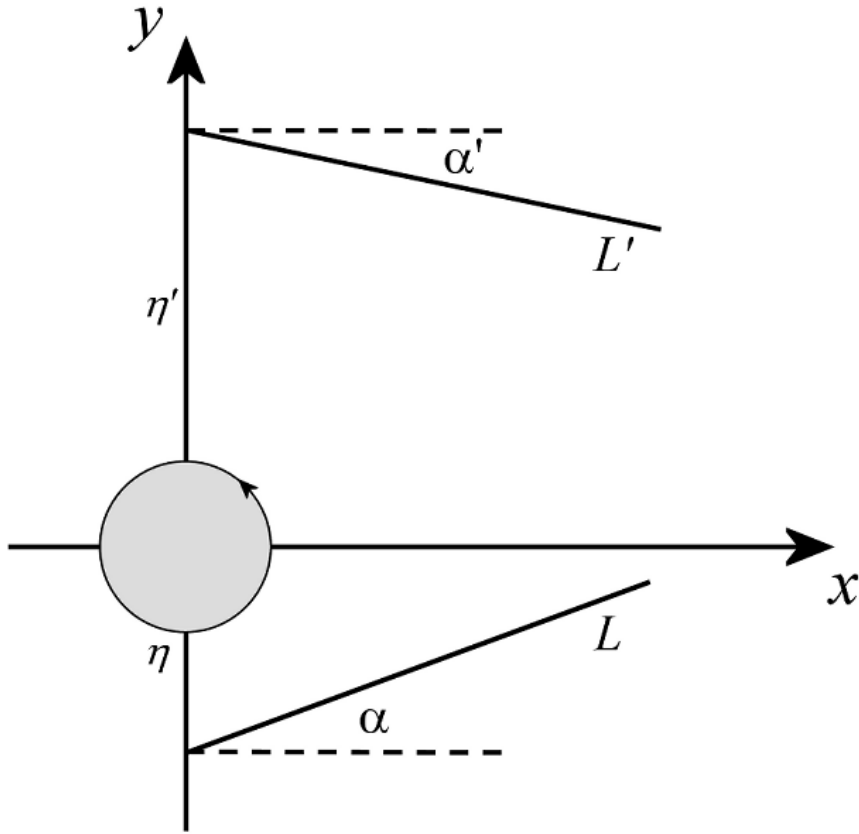}
\end{center}
\caption{The equatorial plane $z=0$ with $a>0$ and thus the
angular momentum points in the positive $z$-direction}
\end{figure}

The calculations are simplest when the particle is projected into
the equatorial plane of the hole. For a Kerr black hole of mass
$m$ and angular momentum ${\rm J}=(0, 0, m\,a)$, with respect to
asymptotically Cartesian spatial coordinates, Figure 1 shows the
asymptotes $L$ and $L'$ of two scattered \emph{photon} paths in
the equatorial plane, one co-rotating with the hole $(L)$ and one
counter-rotating with the hole $(L')$. The small angles of
scattering $\alpha$ and $\alpha '$ indicated in the figure have
been calculated by Boyer and Lindquist \cite{BL} and are found to
be
\begin{equation}\label{2.1}
\alpha =\frac{4\,m}{\eta}-\frac{4\,m\,a}{\eta ^2}>0\
,\end{equation} and
\begin{equation}\label{2.2}
\alpha '=-\frac{4\,m}{\eta '}-\frac{4\,m\,a}{\eta '^2}<0\
,\end{equation} respectively, where $\eta >0$ and $\eta '>0$ are
large impact parameters. To have $\alpha '=-\alpha$ we see that we
must have $\eta '=\eta +2\,a$. In this case for $x\geq 0$ {\it the
straight lines $L$ and $L'$ intersect on the straight line}
\begin{equation}\label{a}y=a,\qquad z=0\ .\end{equation} This asymmetrical behavior is entirely
due to the rotation of the black hole. If $a=0$ then $\eta '=\eta$
and $L$ and $L'$ will intersect on the straight line $y=0$. The
corresponding angles of deflection of \emph{high speed particles},
calculated using the time--like geodesics of (\ref{1.1}) with
(\ref{1.2}) are given exactly by \cite{CQG}
\begin{equation}\label{2.3}
\tan\alpha =\frac{4\,m\,(\eta +a)}{(\eta +a)^2-4\,m^2}\
,\end{equation}and
\begin{equation}\label{2.4}
\tan\alpha '=-\frac{4\,m\,(\eta '-a)}{(\eta '-a)^2-4\,m^2}\
,\end{equation}respectively. Putting $\alpha '=-\alpha$ we deduce
from these equations that
\begin{equation}\label{2.5}
\eta '=\eta +2\,a\ ,\end{equation}without the assumption that
$\eta ,\ \eta '$ are large and so $L$ and $L'$ intersect on the
straight line $y=a,\ z=0$ in this case also. We note that
(\ref{2.3}) and (\ref{2.4}) agree with (\ref{2.1}) and (\ref{2.2})
if the impact parameters are large. This asymmetrical intersection
behavior can be seen in greater generality by studying the
deflection of a family of high speed particles projected into the
field of the Kerr black hole. This involves the study of
time--like geodesics of the space--time with line--element given
by (\ref{1.1}) and (\ref{1.2}).

The time--like geodesic equations for the space--time described by
(\ref{1.1}) and (\ref{1.2}) have been solved in \cite{CQG}. If
$\tau$ is proper time along them then we shall take $\tau <0$ on a
geodesic before it intersects the null hyperplane $x+t=0$, $\tau
=0$ at the point of intersection with $x+t=0$ and $\tau >0$ to the
future of $x+t=0$. For the simplest initial conditions we find
that a time--like geodesic is given for $\tau <0$ by
\begin{eqnarray}\label{2.6}
x&=&x_0+x_1\tau\ ,\\
y&=&y_0\ ,\\
z&=&z_0\ ,\\
t&=&-x_0+t_1\tau\ ,\end{eqnarray}and for $\tau >0$ by
\begin{eqnarray}\label{2.7}
x&=&x_0+x_1\tau +X_1\tau +\hat X_1\ ,\\
y&=&y_0+Y_1\tau\ ,\\
z&=&z_0+Z_1\tau\ ,\\
t&=&-x_0+t_1\tau -X_1\tau -\hat X_1\ ,\end{eqnarray} with
\begin{eqnarray}\label{2.8}
\hat X_1&=&(h)_0\ ,\\
X_1&=&-\frac{C_0}{2}\,\{(h_y)_0^2+(h_z)_0^2\}\ ,\\
Y_1&=&-C_0\,(h_y)_0\ ,\\
Z_1&=&-C_0\,(h_z)_0\ ,\end{eqnarray} and
\begin{equation}\label{2.9}
x_1+t_1=C_0\ ,\qquad x_1-t_1=-C_0^{-1}\ ,\end{equation} where
$C_0$ is a constant. Here $h(y, z)$ is given by (\ref{1.2}),
subscripts $y$ and $z$ on $h$ indicate partial derivatives and
round brackets followed by a subscript zero indicate that $h$ and
its partial derivatives are to be evaluated at $y=y_0 ,\ z=z_0$.
By varying $x_0, y_0, z_0$ we have here two time--like
congruences, one for $\tau <0$ and another for $\tau >0$. The
interesting one corresponds to $\tau >0$ and the unit time--like
tangent to the congruence on $\tau =0$ has components (in
rectangular Cartesian coordinates and time; here and throughout
Greek indices take values 1, 2, 3, 4)
\begin{equation}\label{2.10}
v^\mu =(x_1+X_1\ ,\ Y_1\ ,Z_1\ ,\ t_1-X_1)\ .\end{equation} We can
consider (2.11)--(2.14) defining $x_0,\ y_0,\ z_0,\ \tau$ as
scalar fields on the region of Minkowskian space--time $\tau\geq
0$. Thus $v^\mu$ becomes a vector field on $\tau\geq 0$ whose
integral curves constitute the congruence of interest. In addition
we can use $x_0,\ y_0,\ z_0,\ \tau$ as coordinates on this region
of space--time, which we label by $X^\mu =(\zeta\ ,\bar\zeta\
,x_0\ ,\tau)$ with $\zeta =y_0+iz_0$. In these coordinates
(\ref{2.10}) is given via the 1--form
\begin{equation}\label{2.11}
v_{\mu}\,dX^\mu =C_0\,dx_0-d\tau\ ,\end{equation}or alternatively
as the covariant vector field \begin{equation}\label{2.12}
v^\mu\,\frac{\partial}{\partial X^\mu}=\frac{\partial}{\partial
\tau}\ .\end{equation}The line--element in the region $\tau\geq 0$
of Minkowskian space--time is given by
\begin{eqnarray}\label{2.13}
ds^2&=&dx^2+dy^2+dz^2-dt^2\ ,\nonumber\\
&=&\left |d\zeta -2\,C_0\,\tau\,\frac{\partial
^2(h)_0}{\partial\bar\zeta ^2}\,d\bar\zeta\right
|^2+2\,C_0\,d\tau\,dx_0-d\tau ^2\ .\end{eqnarray}Before the
time--like congruence encounters the null hyperplane history of
the impulsive gravitational wave (for $\tau <0$) it is given
parametrically by (2.7)--(2.10). Again using $x_0, y_0, z_0, \tau$
as cordinates the line--element of Minkowskian space--time for
$\tau <0$ is given by
\begin{equation}\label{2.14}
ds^2=|d\zeta |^2+2\,C_0\,d\tau\,dx_0-d\tau ^2 ,\end{equation}where
as before $\zeta =y_0+iz_0$. The two line--elements (\ref{2.13})
and (\ref{2.14}) can be combined to read
\begin{equation}\label{2.15}
ds^2=\left |d\zeta -2\,C_0\,\tau\,\vartheta (\tau
)\,\frac{\partial ^2(h)_0}{\partial\bar\zeta ^2}\,d\bar\zeta\right
|^2+2\,C_0\,d\tau\,dx_0-d\tau ^2\ .\end{equation}Here $\vartheta
(\tau )$ is the Heaviside step function (equal to unity if $\tau
>0$ and equal to zero if $\tau <0$) and now this line--element
incorporates the impulsive gravitational wave with history $\tau
=0$. The metric tensor in these coordinates is clearly continuous
across $\tau =0$ but has a jump in its first derivative across
$\tau =0$ and the Riemann curvature tensor will have a delta
function singular on $\tau =0$. This can easily be checked by
direct calculation starting with the line--element (\ref{2.15}).

To work out the properties of the time--like congruence
(contraction, twist and shear) for $\tau >0$, having as tangent
the covariant vector field (\ref{2.11}), we need the covariant
derivative of the tangent with respect to the Riemannian
connection calculated with the metric tensor given via the
line--element (\ref{2.15}) with $\tau >0$. With this derivative
indicated by a stroke, we find that it takes the neat form
\begin{equation}\label{2.16}
v_{\mu |\nu}=\hat\mu _1\,n_{(1)\mu}\,n_{(1)\nu}+\hat\mu
_2\,n_{(2)\mu}\,n_{(2)\nu}=v_{\nu |\mu}\ .\end{equation} Here
$n_{(1)\mu}\ ,n_{(2)\mu}$ are given via the 1--forms
\begin{eqnarray}\label{2.17}
n_{(1)\mu}\,dX^\mu
&=&\frac{i}{2}(1+2\,C_0\,\tau\,|h_{\zeta\zeta}|)\,\left\{\left
(\frac{h_{\zeta\zeta}}{|h_{\zeta\zeta}|}\right
)^{\frac{1}{2}}d\zeta -\left
(\frac{|h_{\zeta\zeta}|}{h_{\zeta\zeta}}\right
)^{\frac{1}{2}}d\bar
\zeta\right\}\ ,\\
n_{(2)\mu}\,dX^\mu
&=&\frac{1}{2}(1-2\,C_0\,\tau\,|h_{\zeta\zeta}|)\,\left\{\left
(\frac{h_{\zeta\zeta}}{|h_{\zeta\zeta}|}\right
)^{\frac{1}{2}}d\zeta +\left
(\frac{|h_{\zeta\zeta}|}{h_{\zeta\zeta}}\right
)^{\frac{1}{2}}d\bar \zeta\right\}\ ,\end{eqnarray} where here and
henceforth we use the shorthand \begin{equation}\label{2.18}
h_{\zeta\zeta}=\frac{\partial ^2(h)_0}{\partial\zeta ^2}\
.\end{equation}The vectors $n^\mu _{(1)}\ ,\ n^\mu _{(2)}$ are
unit space--like vectors, orthogonal to each other and each
orthogonal to $v^{\mu}$. They are both parallel transported along
the congruence tangent to $v^\mu$. We thus confirm from
(\ref{2.16}) that the congruence is geodesic and twist--free. The
scalars $\hat\mu _1,\ \hat\mu _2$ in (\ref{2.16}) are given by
\begin{eqnarray}\label{2.19}
\hat\mu
_1&=&\frac{2\,C_0\,|h_{\zeta\zeta}|}{1+2\,C_0\,\tau\,|h_{\zeta\zeta}|}\
,\\ \hat\mu
_2&=&\frac{-2\,C_0\,|h_{\zeta\zeta}|}{1-2\,C_0\,\tau\,|h_{\zeta\zeta}|}\
.\end{eqnarray}Hence the contraction of the congruence is given by
\begin{equation}\label{2.20}
\vartheta =v^{\mu}{}_{|\mu}=\hat\mu _1+\hat\mu
_2=-\frac{8\,C_0^2\tau\,|h_{\zeta\zeta}|^2}{1-4C_0^2\tau
^2|h_{\zeta\zeta}|^2}\ .\end{equation}The shear tensor $\sigma
_{\mu\nu}$ associated with the congruence reads now
\begin{equation}\label{2.21}
\sigma _{\mu\nu}=v_{\mu
|\nu}-\frac{1}{3}\vartheta\,(g_{\mu\nu}+v_\mu\,v_\nu )\
.\end{equation}An orthonormal tetrad which is parallel propagated
along the congruence is given by $\{n_{(1)}^\mu , n_{(2)}^\mu ,
n^\mu , v^\mu\}$ with the unit space--like vector field $n^\mu$
given by
\begin{equation}\label{2.22}
n^\mu\,\frac{\partial}{\partial
X^\mu}=C_0^{-1}\frac{\partial}{\partial
x_0}+\frac{\partial}{\partial \tau}\ .\end{equation}Using
(\ref{2.16}) and (\ref{2.21}) we can write the shear tensor in the
form
\begin{equation}\label{2.23}
\sigma _{\mu\nu}=\mu _1\,n_{(1)\mu}\,n_{(1)\nu}+\mu
_2\,n_{(2)\mu}\,n_{(2)\nu}+\mu _3\,n_{\mu}\,n_{\nu}\
,\end{equation}with \begin{equation}\label{2.24} \mu _1=\hat\mu
_1-\frac{1}{3}\vartheta\ ,\mu _2=\hat\mu _2-\frac{1}{3}\vartheta\
,\mu _3 =-\frac{1}{3}\vartheta\ .\end{equation}We note that
$\{n_{(1)}^\mu , n_{(2)}^\mu , n^\mu , v^\mu\}$ are the unit
eigenvectors of the shear tensor with $\{\mu _1, \mu _2, \mu _3,
0\}$ the corresponding eigenvalues. The shear tensor is trace free
in the sense that $g^{\mu\nu}\,\sigma _{\mu\nu}=0$ and thus $\mu
_1+\mu _2+\mu _3=0$ and this follows from (\ref{2.20}). This
time--like congruence is an approximation to the world lines of
high speed particles \emph{after} they have been deflected by the
rotating black hole. That it exhibits the intersection behavior
described at the beginning of this section (actually a
generalization of this behavior because the congruence above has
resulted from the scattering of high speed particles projected at
{\it any} angle into the field of the black hole) can be seen
easily by first noting from (\ref{2.20}) that the lines of the
congruence converge when $\tau =\tau _0$ with $\tau _0$ given by
\begin{equation}\label{2.25}
\tau _0=\frac{(y_0-a)^2+(z_0+b)^2}{4\,C_0\,p}\
.\end{equation}However from (2.12) and (2.13) we have
\begin{eqnarray}\label{2.26}
y-a&=&(y_0-a)\,\left (1-\frac{\tau}{\tau _0}\right )\
,\\z+b&=&(z_0+b)\,\left (1-\frac{\tau}{\tau _0}\right )\
,\end{eqnarray}and the right hand sides of these equations vanish
when $\tau =\tau _0$ given by (\ref{2.25}). Thus in this more
general case \emph{the paths of the high speed particles converge
on} a straight line after being scattered by the black hole and
this straight line is \emph{the intersection of the planes} $y=a,\
z=-b$.

\setcounter{equation}{0}
\section{Scattering of Particles:Null Geodesics}\indent
We will consider in section 4 the collision of plane
electromagnetic waves propagating in the positive $x$--direction
with the impulsive gravitational wave (\ref{1.1}) with (\ref{1.2})
which is propagating in the negative $x$--direction. We will treat
the electromagnetic field of the waves before and after collision
as a test field on this space--time. We show in an appendix that
it is always possible to find a frame of reference in which this
collision is head--on. As a prelude to this we examine in this
section the behavior of null geodesics in the space--time
described by (\ref{1.1}) and (\ref{1.2}). The light--like
propagation direction of the electromagnetic waves before
encountering the impulsive gravitational wave is given (in
rectangular Cartesian coordinates and time $x, y, z, t$) by
\begin{equation}\label{3.1}
{}^{(-)}l^{\mu}=(1, 0, 0, 1)\ .\end{equation} The superscript
minus on $l$ denotes a quantity before collision with the
gravitational wave. A useful null tetrad is given by
$\{{}^{(-)}l^{\mu}, {}^{(-)}n^{\mu}, {}^{(-)}m^{\mu}, {}^{(-)}\bar
m^{\mu}\}$ with
\begin{equation}\label{3.2}
{}^{(-)}n^{\mu}=\left(\frac{1}{2}, 0, 0,-\frac{1}{2}\right )\
,\qquad {}^{(-)}m^{\mu}=\frac{1}{\sqrt{2}}(0, 1, i, 0)\
,\end{equation} and ${}^{(-)}\bar m^{\mu}$ the complex conjugate
of ${}^{(-)}m^{\mu}$. All scalar products among
$\{{}^{(-)}l^{\mu}, {}^{(-)}n^{\mu}, {}^{(-)}m^{\mu}, {}^{(-)}\bar
m^{\mu}\}$ vanish with the exception of
${}^{(-)}n^{\mu}\,{}^{(-)}l_{\mu}=1$ and ${}^{(-)}\bar
m^{\mu}\,{}^{(-)}m_{\mu}=1$. As a preliminary we investigate what
happens a null geodesic starting with tangent ${}^{(-)}l^{\mu}$ in
the space--time with line--element (\ref{1.1}). The differential
equations of such a curve $x^{\mu}=x^{\mu}(\lambda )$, where
$\lambda$ is an affine parameter, are
\begin{eqnarray}\label{3.3}
\ddot x&=&H_x\,(\dot x+\dot t)^2+2\,(h_y\dot y+h_z\dot z)\,(\dot
x+\dot t)\,\delta
(x+t)\ ,\\
\ddot y&=&-h_y\,(\dot x+\dot t)^2\delta (x+t)\ ,\\
\ddot z&=&-h_z\,(\dot x+\dot t)^2\delta (x+t)\ ,\\
\ddot t&=&-H_x\,(\dot x+\dot t)^2-2\,(h_y\dot y+h_z\dot z)\,(\dot
x+\dot t)\,\delta (x+t)\ .\end{eqnarray}As always subscripts on
$H$ and $h$ denote partial derivatives and, in particular,
$H_x=h(y, z)\,\delta '(x+t)$, with the prime on the delta function
denoting the derivative with respect to the argument of the
function. Differentiation with respect to $\lambda$ is indicated
by a dot. The first integral of (3.3)--(3.6) is given by
\begin{equation}\label{3.4}
(\dot x+\dot t)\,\{\dot x-\dot t-2\,H\,(\dot x+\dot t)\}+\dot
y^2+\dot z^2=0\ .\end{equation} These equations are identical in
form to the time--like geodesic equations in this space--time
considered in \cite{CQG}, with the exception of a zero appearing
on the right hand side of (\ref{3.4}) instead of minus one. They
are solved in exactly the same way. If $\lambda <0$ on the null
geodesic before encountering the null hyperplane
$x+t=0\Leftrightarrow\lambda =0$, which is the history of the
impulsive gravitational wave, and $\lambda
>0$ on the null geodesic after encountering the null hyperplane $x+t=0$ then we find
that for $\lambda <0$ the geodesic is given by
\begin{equation}\label{3.5}
x=x_0+\lambda\ ,\qquad y=y_0\ ,\qquad z=z_0\ ,\qquad
t=-x_0+\lambda\ ,\end{equation} where $x_0, y_0, z_0$ are
constants of integration. For $\lambda >0$ the geodesic is given
by
\begin{eqnarray}\label{3.6}
x&=&x_0+\lambda +X_1\,\lambda +\hat X_1\ ,\\
y&=&y_0+Y_1\,\lambda\ ,\\ z&=&z_0+Z_1\,\lambda\
,\\
t&=&-x_0+\lambda -X_1\,\lambda -\hat X_1\ .\end{eqnarray} Here
\begin{equation}\label{3.7}
\hat X_1=(h)_0\ ,\qquad Y_1=-2\,(h_y)_0\ ,\qquad Z_1=-2\,(h_z)_0\
,\end{equation} and
\begin{equation}\label{3.8}
X_1=-(h_y)_0^2-(h_z)_0^2\ ,\end{equation} with, as before, the
round brackets followed by a subscript zero on a function of $y$
and $z$ denoting that the function is calculated at $y=y_0,\
z=z_0$ (on $\lambda =0$). We see from (3.8)--(3.12) that the point
at which the null geodesic meets the null hyperplane and the point
from which it leaves the null hyperplane (they are different
points due to a translation along the generators of $\lambda =0$
in going from the past side of the null hyperplane to the future
side, in order to construct the impulsive wave geometrically) are
both specified by the constants $x_0, y_0, z_0$. We note that the
tangent to the null geodesic on the future side of $\lambda =0$ is
given by
\begin{equation}\label{3.9}
l^{\mu}=(1+X_1, Y_1, Z_1, 1-X_1) .\end{equation}A null tetrad
defined on the future side of $\lambda =0$ at the point specified
by $x_0, y_0, z_0$ is given by $\{l^{\mu}, n^{\mu}, m^{\mu}, \bar
m^{\mu}\}$ with
\begin{equation}\label{3.10}
n^{\mu}=\left (\frac{1}{2}, 0, 0,-\frac{1}{2}\right
)={}^{(-)}n^\mu\ ,\end{equation}and
\begin{equation}\label{3.11}
m^{\mu}=\frac{1}{\sqrt{2}}\left(-\frac{1}{2}\left(Y_1+iZ_1\right
), 1, i, \frac{1}{2}\left(Y_1+iZ_1\right )\right )\
,\end{equation} with $\bar m^{\mu}$ the complex conjugate of
$m^{\mu}$.

The vectors $\{l^{\mu}, n^{\mu}, m^{\mu}, \bar m^{\mu}\}$ defined
on the future side of $\lambda =0$ in (\ref{3.9})--(\ref{3.11})
can be extended to vector fields in the region to the future of
the null hyperplane (the region corresponding to $\lambda >0$) by
parallel transport along the null geodesics tangent to $l^{\mu}$
emanating to the future from each point of  $\lambda =0$ specified
by $x_0, y_0, z_0$. We then need to calculate the derivatives of
these vector fields with respect to $x, y, z$ and $t$. To do this
we notice first that $x_0, y_0, z_0$ and $\lambda$ can be extended
to scalar fields for $\lambda
>0$ by reading (3.9)--(3.12) as defining $x_0, y_0, z_0, \lambda$ as
functions of $x, y, z, t$. Now the derivatives of $x_0, y_0, z_0,
\lambda$ each with respect to $x, y, z, t$ can be calculated from
these equations. Among these derivatives are the following with a
transparent geometrical interpretation:
\begin{equation}\label{3.12}
\frac{\partial\lambda}{\partial x^{\mu}}=n_{\mu}\qquad {\rm
and}\qquad \frac{\partial x_0}{\partial
x^{\mu}}=\frac{1}{2}\,l_{\mu}\ .\end{equation} Thus the integral
curves of the vector fields $n^{\mu},\ l^{\mu}$ are twist--free
null geodesics. For the case of $n^{\mu}$ they generate the null
hyper{\it planes} $\ \lambda ={\rm constant}$ while for $l^{\mu}$
they generate the null hyper{\it surfaces} $\ x_0={\rm constant}$.
Clearly $n^{\mu}$ is a constant vector field (in the rectangular
Cartesian coordinates and time) and so this property of the
integral curves of $n^{\mu}$ is obvious. This is not the case for
$l^{\mu}$ and we will confirm below that the integral curves of
$l^{\mu}$ have contraction and shear. The derivatives of $y_0$ and
$z_0$ are given as follows: Defining
\begin{equation}\label{3.13}
\varphi =1-4\,\lambda ^2\{(h_{yz})^2_0+(h_{yy})^2_0\}\
,\end{equation} we find that
\begin{eqnarray}\label{3.14}
\varphi\,\frac{\partial y_0}{\partial
x}&=&-\frac{1}{2}\,\left\{(y_1+Y_1)\,\left
(1+\lambda\,\frac{\partial Z_1}{\partial z_0}\right
)-(z_1+Z_1)\,\lambda\,\frac{\partial Y_1}{\partial
z_0}\right\}\nonumber\\&=&\varphi\,\frac{\partial y_0}{\partial t}\ ,\\
\varphi\,\frac{\partial z_0}{\partial
x}&=&\frac{1}{2}\,\left\{-(z_1+Z_1)\,\left
(1+\lambda\,\frac{\partial Y_1}{\partial y_0}\right
)+(y_1+Y_1)\,\lambda\,\frac{\partial Z_1}{\partial
y_0}\right\}\nonumber\\&=&\varphi\,\frac{\partial z_0}{\partial t}\ ,\\
\varphi\,\frac{\partial y_0}{\partial
y}&=&1+\lambda\,\frac{\partial Z_1}{\partial z_0}\ ,\qquad
\varphi\,\frac{\partial z_0}{\partial
y}=-\lambda\,\frac{\partial Z_1}{\partial y_0}\ ,\\
\varphi\,\frac{\partial y_0}{\partial
z}&=&-\lambda\,\frac{\partial Y_1}{\partial z_0}\
,\qquad\varphi\,\frac{\partial z_0}{\partial
z}=1+\lambda\,\frac{\partial Y_1}{\partial y_0}\ ,\end{eqnarray}
Lowering indices with the Minkowskian metric tensor in rectangular
Cartesian coordinates and time $\eta _{\mu\nu}={\rm diag}(1, 1, 1,
-1)$ and denoting partial derivatives by a comma we obtain from
these
\begin{equation}\label{3.15} \sigma =l_{\mu ,\nu}\,m^\mu\,m^\nu =
\frac{-4\,\{(h_{yy})_0+i(h_{yz})_0\}}{1-4\,\lambda
^2\{(h_{yy})_0^2+(h_{yz})_0^2\}}\ ,\end{equation} and
\begin{equation}\label{3.16}
\rho =l_{\mu ,\nu}\,m^\mu\,\bar m^\nu
=\frac{-4\,\lambda\,\{(h_{yy})_0^2+(h_{yz})_0^2\}}{1-4\,\lambda
^2\{(h_{yy})_0^2+(h_{yz})_0^2\}}\ .\end{equation}The scalar
$\sigma$ is the (complex) shear of the null geodesic congruence
tangent to $l^{\mu}$ while the real scalar $\rho$ is the
contraction of this congruence. Since ${}^{(-)}l^{\mu}$ introduced
in (\ref{3.1}) is a constant vector field its integral curves are
twist--free, expansion--free, shear--free null geodesics. We see
now that on crossing $x+t=0\Leftrightarrow\lambda =0$ this
congruence experiences a jump in the shear while the expansion is
continuous. These are general properties of a null geodesic
congruence crossing the history of an impulsive gravitational wave
first pointed out by Penrose \cite{P}. The former of the two
properties exhibits the wave behaving as an astigmatic lens while
the latter property demonstrates the focussing effect of this
`lens'. The plane electromagnetic waves incident on the high speed
black hole will have propagation direction ${}^{(-)}l^{\mu}$ in
space--time before colliding with the impulsive gravitational
wave. Since this propagation direction acquires contraction after
encountering the gravitational wave the electromagnetic waves
cannot remain plane and, in fact, since this direction acquires
shear it cannot remain the propagation direction of pure
electromagnetic waves (except perhaps in an approximation) on
account of Robinson's theorem \cite{R}.

With the derivatives of $x_0,\ y_0,\ z_0,\ \lambda$ with respect
to $x, y, z, t$ now known the derivatives of $X_1, Y_1, Z_1$ in
(\ref{3.7}) and (\ref{3.8}) can be evaluated and using these the
derivatives of the vector fields $l^\mu ,\ n^\mu ,\ m^\mu $ with
respect to $x, y, z, t$ can be derived. Lowering indices with the
Minkowskian metric tensor in rectangular Cartesian coordinates and
time $\eta _{\mu\nu}={\rm diag}(1, 1, 1, -1)$ and denoting partial
derivatives by a comma, we obtain the deceptively simple results:
\begin{eqnarray}\label{3.17}
l_{\mu ,\nu}&=&\bar\sigma\,m_{\mu}\,m_{\nu}+\sigma\,\bar
m_{\mu}\,\bar m_{\nu}+\rho\,(m_{\mu}\,\bar m_{\nu}+\bar
m_{\mu}\,m_{\nu})\ ,\\
n_{\mu ,\nu}&=&0\ ,\\
m_{\mu ,\nu}&=&-\rho\,n_{\mu}\,m_{\nu}-\sigma\,n_{\mu}\,\bar
m_{\nu}\ ,\end{eqnarray}with $\sigma$ and $\rho$ given by
(\ref{3.15}) and (\ref{3.16}).

We see from (\ref{3.16}) that $\rho$ becomes infinite
(neighbouring null geodesics tangent to $l^\mu$ intersect) when
$\lambda =\lambda _0>0$ given by
\begin{equation}\label{3.17}
4\,\lambda _0^2\{(h_{yy})_0^2+(h_{yz})_0^2\}=1\ .\end{equation}
Substituting for $h(y, z)$ from (\ref{1.2}) we see that this
occurs when
\begin{equation}\label{3.18}
8\,p\,\lambda _0=(y_0-a)^2+(z_0+b)^2\ .\end{equation} On account
of (3.10) and (3.11) we have in general for $\lambda >0$,
\begin{eqnarray}\label{3.19}
y-a&=&(y_0-a)\,\left (1-\frac{\lambda}{\lambda
 _0}\right )\ ,\\z+b&=&(z_0+b)\,\left (1-\frac{\lambda}{\lambda _0}\right
)\ .\end{eqnarray}Hence for $\lambda =\lambda _0$ given in
(\ref{3.18}) we see that the integral curves of $l^\mu$ focus on
the straight line \begin{equation}\label{3.20} y=a\ ,\qquad z=-b\
.\end{equation}This behavior of the null geodesics mirrors the
behavior of the time--like geodesics described in the previous
section. The fact that the rays converge on a straight line shows
that the lensing source is moving on a straight line.

In the region of space--time corresponding to $\lambda
>0$ we may regard $x_0, y_0, z_0, \lambda$ , in the
terminology of Synge \cite{S}, as optical coordinates relative to
a null hyperplane base $\lambda =0$. The coordinate transformation
relating these coordinates to the rectangular Cartesians and time
is given by (3.9)--(3.12). Putting $\zeta =y_0+iz_0$ we find that
the line--element of Minkowskian space--time in coordinates
$X^{\mu}=(\zeta ,\bar\zeta , \lambda , x_0)$ reads
\begin{equation}\label{3.20}
ds^2=\left |d\zeta +2\,\lambda\,\frac{\partial
Y_1}{\partial\bar\zeta}\,d\bar\zeta\right |^2+4\,d\lambda\,dx_0\
.\end{equation} Denoting the null tetrad $l^\mu ,\ n^\mu ,\ m^\mu
,\ \bar m^\mu$ defined in (\ref{3.9})--(\ref{3.11}) by $L^{\mu}$,
$\ N^{\mu}$, $\ M^{\mu}$, $\bar M^{\mu}$ when expressed in the
coordinates $X^\mu$, we find that it is given via the 1--forms
(cf. the equations in ({\ref{3.12}))
\begin{eqnarray}\label{3.21}
L_{\mu}\,dX^{\mu}&=&2\,dx_0\ ,\\
N_{\mu}\,dX^{\mu}&=&d\lambda\ ,\\
M_{\mu}\,dX^{\mu}&=&\frac{1}{\sqrt{2}}\left (d\zeta +2\,\lambda\,
\frac{\partial Y_1}{\partial\bar\zeta}\,d\bar\zeta\right )\ ,\\
\bar M_{\mu}\,dX^{\mu}&=&\frac{1}{\sqrt{2}}\left (d\bar\zeta
+2\,\lambda\, \frac{\partial Y_1}{\partial\zeta}\,d\zeta\right )\
.\end{eqnarray} In the coordinates $X^{\mu}$ the important
equations (3.26)--(3.28) are now written as
\begin{eqnarray}\label{3.22}
L_{\mu |\nu}&=&\bar\sigma\,M_{\mu}\,M_{\nu}+\sigma\,\bar
M_{\mu}\,\bar M_{\nu}+\rho\,(M_{\mu}\,\bar M_{\nu}+\bar
M_{\mu}\,M_{\nu})\ ,\\
N_{\mu |\nu}&=&0\ ,\\
M_{\mu |\nu}&=&-\rho\,N_{\mu}\,M_{\nu}-\sigma\,N_{\mu}\,\bar
M_{\nu}\ ,\end{eqnarray}where the stroke indicates covariant
differentiation with respect to the Riemannian connection
associated with the metric tensor given via the line--element
(\ref{3.20}). The scalars $\sigma$ and $\rho$ are still given by
(\ref{3.15}) and (\ref{3.16}) respectively which can also be
written in the form:
\begin{eqnarray}\label{3.23}
\sigma &=&2\,\frac{\partial Y_1}{\partial\bar\zeta}\,\left
(1-4\,\lambda ^2\left |\frac{
\partial Y_1}{\partial\zeta}\right |^2\right )^{-1}\ ,\\
\rho &=&-4\,\lambda \left |\frac{\partial
Y_1}{\partial\zeta}\right |^2\left (1-4\,\lambda ^2\left |\frac{
\partial Y_1}{\partial\zeta}\right |^2\right )^{-1}\ .\end{eqnarray}
With $Y_1$ in (\ref{3.7}) and $h$ in (\ref{1.2}) we can now write
\begin{equation}\label{3.24}
\frac{\partial Y_1}{\partial\zeta}=\frac{4\,p}{(\zeta -\alpha
)^2}\ ,\end{equation} with $\alpha =a-ib$. We see from (\ref{3.5})
that the coordinates $\zeta , \bar\zeta , \lambda , x_0$ are also
available to us in the region $\lambda <0$. In this case the
Minkowskian line--element takes the form
\begin{equation}\label{3.25}
ds^2=|d\zeta |^2+4\,d\lambda\,dx_0\ .\end{equation} We now have a
null tetrad basis in $\lambda <0$ given via the 1--forms
\begin{equation}\label{3.26}
{}^{(-)}L_{\mu}\,dX^{\mu}=2\,dx_0\ ,\qquad
{}^{(-)}N_{\mu}\,dX^{\mu}=d\lambda\ ,
\end{equation}and \begin{equation}
\label{3.27}{}^{(-)}M_{\mu}\,dX^{\mu}=\frac{1}{\sqrt{2}}\,d\zeta\
, \qquad {}^{(-)}\bar
M_{\mu}\,dX^{\mu}=\frac{1}{\sqrt{2}}\,d\bar\zeta \ .\end{equation}
We note that {\it on} $\lambda =0$ we have
${}^{(-)}L^{\mu}=L^{\mu} ,\ {}^{(-)}N^{\mu}=N^{\mu} ,\
{}^{(-)}M^{\mu}=M^{\mu}$ and ${}^{(-)}\bar M^{\mu}=\bar M^{\mu}$.
Making use of the Heaviside step function $\vartheta (\lambda )$
we can incorporate (\ref{3.20}) and (\ref{3.25}) in the one
expression,
\begin{equation}\label{3.28}
ds^2=\left |d\zeta +2\,\lambda\,\vartheta
(\lambda)\,\frac{\partial
Y_1}{\partial\bar\zeta}\,d\bar\zeta\right |^2+4\,d\lambda\,dx_0\
.\end{equation}This form of the line--element also includes the
impulsive gravitational wave.

\setcounter{equation}{0}
\section{Scattering of Electromagnetic Waves: Dimming of the Signal}\indent
To study the scattering of plane electromagnetic waves by a high
speed Schwarzschild black hole (the Kerr case is qualitatively the
same as this and is mentioned following (\ref{4.21}) below) we
treat the Maxwell fields before and after encountering the
impulsive gravitational wave as test fields in the space--time
with line--element (\ref{1.1}) with $a=b=0$. Thus the incoming
plane waves will have ${}^{(-)}L^{\mu}$ as propagation direction
in space--time for $\lambda <0$. If $F^{\mu\nu}=-F^{\nu\mu}$ is
the Maxwell bivector and $F^*_{\mu\nu}=\frac{1}{2}\eta
_{\mu\nu\rho\sigma}\,F^{\rho\sigma}$ is its dual (here $\eta
_{\mu\nu\rho\sigma}=\sqrt{-g}\,\epsilon _{\mu\nu\rho\sigma}$ where
$g={\rm det}(g_{\mu\nu})$ and $\epsilon _{\mu\nu\rho\sigma}$ is
the Levi--Civita permutation symbol) and if ${\cal
F}_{\mu\nu}=F_{\mu\nu}+iF^*_{\mu\nu}$ then the plane wave field
has the algebraic form
\begin{equation}\label{4.1}
{}^{(-)}{\cal
F}^{\mu\nu}={}^{(-)}A\,\{{}^{(-)}L^{\mu}\,{}^{(-)}M^{\nu}-{}^{(-)}L^{\nu}\,{}^{(-)}M^{\mu}\}\
,
\end{equation}where ${}^{(-)}A$ is a complex--valued function of the coordinates
$\{\zeta , \bar\zeta , \lambda , x_0\}$. Maxwell's source--free
field equations in the region $\lambda <0$ read
\begin{equation}\label{4.2}
{}^{(-)}{\cal F}^{\mu\nu}{}_{,\nu}=0 .\end{equation}By
(\ref{3.26}) and (\ref{3.27}) these yield
\begin{equation}\label{4.3}
{}^{(-)}A={}^{(-)}A(x_0)\ .\end{equation} We see that ${}^{(-)}A$
is constant on the history of a plane wave front (the null
hyperplane $x_0={\rm constant}$) and, of course, we are free to
specify the profile of the incoming waves. In the region $\lambda
>0$ of space--time after the collision with the impulsive
gravitational wave the algebraic form of the test Maxwell field is
completely general:
\begin{eqnarray}\label{4.4}
{}^{(+)}{\cal
F}^{\mu\nu}&=&A\,(L^{\mu}\,M^{\nu}-L^{\nu}\,M^{\mu})+B\,(N^{\mu}\,\bar
M^{\nu}
-N^{\nu}\,\bar M^{\mu})\nonumber\\
&&+C\,(M^{\mu}\,\bar M^{\nu}-M^{\nu}\,\bar
M^{\mu}+L^{\mu}\,N^{\nu}-L^{\nu}\,N^{\mu})\ ,
\end{eqnarray}with $A, B, C$ complex--valued functions of the coordinates
$\{\zeta , \bar\zeta , \lambda , x_0\}$. The source--free Maxwell
field equations in the region $\lambda >0$ read
\begin{equation}\label{4.5}
{}^{(+)}{\cal F}^{\mu\nu}{}_{|\nu}=0 ,\end{equation}and these are
calculated with the help of (3.35)--(3.38). We thus arrive at the
propagation equations along $L^\mu$,
\begin{eqnarray}\label{4.6}
\frac{\partial A}{\partial\lambda}+\rho\,A&=&C_{,\mu}\bar M^\mu\ ,\\
\frac{\partial C}{\partial\lambda}+2\,\rho\,C&=&B_{,\mu}\bar
M^\mu\ ,
\end{eqnarray}
and the `constraint' equations
\begin{eqnarray}\label{4.7}
A_{,\mu}\,M^\mu +\frac{1}{2}\,\frac{\partial C}{\partial x_0}&=&0\ ,\\
C_{,\mu}\,M^\mu +\frac{1}{2}\,\frac{\partial B}{\partial
x_0}&=&\sigma\,A\ ,\end{eqnarray} with $\rho$ and $\sigma$ given
by (3.43), (3.44) and (\ref{3.24}). In (\ref{4.4}) the $A$--part
is Type N in the Petrov classification with degenerate principal
null direction $L^\mu$ and thus it may be considered as describing
`transmitted waves', the $B$--part is also Type N but with $N^\mu$
as degenerate principal null direction and since
$N^{\mu}\,L_{\mu}=1$ we may consider the $B$-- part as
representing `backscattered waves'. The $C$--part of (\ref{4.4})
is algebraically general with principal null directions $L^{\mu},\
N^{\mu}$ and thus is a `Coulomb part' of the electromagnetic
field. Clearly from (4.9) we cannot have, when $\lambda >0$,
$B=C=0$ because $\sigma\neq 0$. In addition with the boundary
conditions given below in (\ref{4.11}) we cannot separately have
$B=0$ or $C=0$ for $\lambda >0$. This is because integrating (4.7)
when $B=0$ would imply $C=0$ for $\lambda >0$ since $C=0$ on
$\lambda =0$ by (\ref{4.11}). Also integrating (4.6) when $C=0$
gives a result inconsistent with (4.8). The source--free Maxwell
test field on the space--time with line--element (\ref{3.28}) can
be written, for any value of $\lambda$, in the form
\begin{equation}\label{4.8}
{\cal F}^{\mu\nu}={}^{(-)}{\cal F}^{\mu\nu}\,(1-\vartheta (\lambda
))+{}^{(+)} {\cal F}^{\mu\nu}\,\vartheta (\lambda )\ .
\end{equation}The Christoffel symbols calculated with the metric given via the line--element (\ref{3.28})
take the form
\begin{equation}\label{4.9}
\Gamma ^{\mu}_{\nu\sigma}={}^{(+)}\Gamma
^{\mu}_{\nu\sigma}\,\vartheta (\lambda )\ .\end{equation} Hence
Maxwell's source--free equations, ${\cal F}^{\mu\nu}{}_{|\nu}=0$
(with the stroke here representing covariant differentiation with
respect to the connection (\ref{4.9})), yield (\ref{4.2}) for
$\lambda <0$, (\ref{4.5}) for $\lambda >0$ and to avoid a
distribution of 4--current on the null hyperplane $\lambda =0$ we
must have, {\it on} $\lambda =0$, the junction conditions:
\begin{equation}\label{4.10}
\left\{{}^{(+)}{\cal F}^{\mu\nu}-{}^{(-)}{\cal
F}^{\mu\nu}\right\}\,N_{\nu}=0\ .\end{equation}Written out
explicitly these equations read, on $\lambda =0$,
\begin{equation}\label{4.11}
A={}^{(-)}A(x_0)\qquad {\rm and}\qquad C=0\ .\end{equation}

We can now evaluate the Maxwell equations (4.6)--(4.9) on $\lambda
=0$ using (\ref{4.11}) to obtain
\begin{equation}\label{4.12}
\frac{\partial A}{\partial\lambda}=0\ ,\qquad \frac{\partial
C}{\partial\lambda}= \sqrt{2}\,\frac{\partial B}{\partial\zeta}\
,\qquad \frac{1}{2}\frac{\partial B}{\partial x_0}
=2\,{}^{(-)}A(x_0)\,\frac{\partial Y_1}{\partial\bar\zeta}\
.\end{equation} We emphasize that these equations only hold {\it
on} $\lambda =0$. The Maxwell equation (4.8) gives $0=0$ in this
case. Writing ${}^{(-)}A(x_0)=d{}^{(-)}W(x_0)/dx_0$ for
convenience, and using (\ref{3.24}), the final equation in
(\ref{3.12}) gives
\begin{equation}\label{4.13}
B=\frac{16\,p\,{}^{(-)}W(x_0)}{\bar\zeta ^2}\ ,
\end{equation}on $\lambda =0$, where a function of integration (of $\zeta\ ,\bar\zeta$) has been
put equal to zero on the basis that, without loss of generality,
if there is no in--coming Maxwell field (${}^{(-)}A(x_0)=0$) then
there is no out--going Maxwell field. We thus see that on $\lambda
=0$, ${}^{(+)}B$ is singular at $\zeta =0$ and tends to zero like
$|\zeta |^{-2}$ for $|\zeta |$ large. As a result of (\ref{4.13})
the second equation in (\ref{4.12}) now reads
\begin{equation}\label{4.14}
\frac{\partial C}{\partial\lambda}=0\ ,
\end{equation}on $\lambda =0$. If we now return to (\ref{4.4}) and define the
separate parts by
\begin{eqnarray}\label{4.15}
{}^{(+)}{\cal F}^{\mu\nu}_1&=&A(L^\mu\,M^\nu -L^\nu\,M^\mu )\ ,\\
{}^{(+)}{\cal F}^{\mu\nu}_2&=&B(N^\mu\,\bar M^\nu -N^\nu\,\bar M^\mu )\ ,\\
{}^{(+)}{\cal F}^{\mu\nu}_3&=&C(M^\mu\,\bar M^\nu -M^\nu\,\bar
M^\mu + L^\mu\,N^\nu -L^\nu\,N^\mu )\ ,\end{eqnarray}then using
(\ref{4.12})--(\ref{4.14}), and the formulae (3.39)--(3.41), we
find that {\it on} $\lambda =0$,
\begin{eqnarray}\label{4.16}
{}^{(+)}{\cal
F}^{\mu\nu}_1{}_{|\nu}&=&\frac{8\sqrt{2}\,p\,{}^{(-)}A(x_0)}{\bar\zeta
^2}\,\delta ^{\mu}_1\ ,\\
{}^{(+)}{\cal
F}^{\mu\nu}_2{}_{|\nu}&=&-\frac{8\sqrt{2}\,p\,{}^{(-)}A(x_0)}{\bar\zeta
^2}\,\delta ^{\mu}_1\ ,\\
{}^{(+)}{\cal F}^{\mu\nu}_3{}_{|\nu}&=&0\ .\end{eqnarray}These
equations allow us to see how each of the complex, self--dual
bivectors above approximates a Maxwell field on $\lambda =0$ for
large values of $|\zeta |$.

In coordinates $\{x, y, z, t\}$ the 4--momentum of the in--coming
photons is given by (\ref{3.1}) while the 4--momentum of the
photons immediately after the collision (i.e. on the future side
of the null hyperplane $\lambda =0$) is given by (\ref{3.9}) which
we write more explicitly as
\begin{equation}\label{4.17}
l^{\mu}=\left (1-\frac{16\,p^2}{R_0^2}, -\frac{8\,p\,\cos\phi
_0}{R_0}, -\frac{8\,p\,\sin\phi _0}{R_0},
1+\frac{16\,p^2}{R_0^2}\right )\ ,\end{equation} and where for
convenience we have written
\begin{equation}\label{4.18}
\zeta =R_0\,{\rm e}^{i\phi}\ ,\end{equation}with $R_0,\ \phi$
real. If
\begin{equation}\label{4.19}
R_0<4\,p\ ,\end{equation} then (\ref{4.17}) is given approximately
by
\begin{equation}\label{4.20}
l^{\mu}=-\frac{16\,p^2}{R_0^2}\,\left (1, 0, 0, -1\right
)=-\frac{32\,p^2}{R_0^2}\,n^{\mu}\ .\end{equation} We note that,
as defined in (\ref{3.10}), $n^\mu$ is past--pointing. We see from
(\ref{4.20}) that photons colliding with the gravitational wave
sufficiently close to the singular point on the wave ($y_0=0,\
z_0=0$) are reflected back in the direction from whence they came
and accompany the gravitational wave moving in the reflected
direction. Hence there are observers, who in the normal course of
events would see the light after it has passed through the
gravitational wave, who can also see a circular disk on their sky.
The equation of this disk is
\begin{equation}\label{4.21}
y_0^2+z_0^2\leq 16\,p^2\ .\end{equation} As this result is only
approximate the disk is not opaque to these observers but the
light passing through it is {\it dimmed} because most of it is
reflected. The case of plane electromagnetic waves scattered by a
high speed Kerr black hole is qualitatively the same as this case.
The only difference is that the singularity on the gravitational
wave front is at $y_0=a,\ z_0=-b$ now and thus the equation of the
disk (\ref{4.21}) is adjusted to read
\begin{equation}\label{4.22}
(y_0-a)^2+(z_0+b)^2\leq 16\,p^2\ .\end{equation}

The electromagnetic energy tensor on the plus (future) side of the
history of the impulsive gravitational wave is now obtained from
\begin{equation}\label{4.22}
2\,{}^{(+)}T^{\mu\sigma}=|{}^{(-)}A|^2L^\mu\,L^\sigma
+|B|^2N^\mu\,N^\sigma +{}^{(-)}\bar A\,B\,\bar M^\mu\,\bar
M^\sigma +{}^{(-)}A\,\bar B\,M^\mu\,M^\sigma\ ,\end{equation}while
the electromagnetic energy tensor on the past side of the history
of the impulsive gravitational wave is given by
\begin{equation}\label{4.23}
2\,{}^{(-)}T^{\mu\sigma}=|{}^{(-)}A|^2L^\mu\,L^\sigma \
.\end{equation}Using (\ref{4.18}) we see that for $R_0$ very large
(\ref{4.22}) is dominated by the first term. Thus, from the point
of view of the electromagnetic energy tensor, the electromagnetic
waves which collide with the impulsive gravitational wave
effectively ``pass through" the gravitational wave in the region
of the gravitational wave front which is far from the singularity
$R_0=0$. For $R_0$ small the situation is quite different. In
general we have
\begin{equation}\label{4.24}
\frac{|B|}{|{}^{(-)}A|}=\frac{16\,p}{R^2_0}\,\left
|\frac{{}^{(-)}W}{{}^{(-)}A}\right |\ .\end{equation}We expect the
contribution to (\ref{4.22}) from the reflected wave, described by
$B$, to be weaker than the incident wave contribution described by
${}^{(-)}A$, and so we have
\begin{equation}\label{4.25}
\frac{|B|}{|{}^{(-)}A|}<1\ .\end{equation}If (\ref{4.19}) holds
then it is particularly interesting to consider incoming waves of
a definite frequency $\omega$ so that
${}^{(-)}A={}^{(-)}A_0\exp\{i\omega\,x_0\}$ where ${}^{(-)}A_0\ ,\
\omega$ are constants. Now we may take
${}^{(-)}W={}^{(-)}A/i\omega$ and (\ref{4.24}) becomes
\begin{equation}\label{4.26}
\frac{|B|}{|{}^{(-)}A|}>\frac{1}{p\,\omega}\ .\end{equation}This
inequality ensures that $B\neq 0$ and so there is a reflected wave
on $\lambda =0$. Now (\ref{4.25}) implies
\begin{equation}\label{4.27}
\omega >\frac{1}{p}\ .\end{equation}This means that for $R_0$
small, in the sense of (\ref{4.19}), \emph{only waves of
sufficiently high frequency will be reflected and will thus fail
to pass through the gravitational wave}. This leads to the dimming
of the signal.

\setcounter{equation}{0}
\section{Discussion}\indent
The singularity in the line--element (\ref{1.1}) with (\ref{1.2})
at $y=a,\ z=-b$ has its origin in the fact that the source of the
Kerr gravitational field is isolated and rotating. We have
demonstrated in section 2 that \emph{the focussing behavior} of a
time--like congruence in the space--time with line--element
(\ref{1.1}) is exactly what one would expect in the field of a
rotating source. In section 3 we calculated the behavior of
photons (a null geodesic congruence) which experience a head--on
collision with the impulsive gravitational wave described by
(\ref{1.1}) and (\ref{1.2}) as a prelude to the study of the
collision of plane electromagnetic waves with the gravitational
wave. This enabled us to demonstrate in section 4 \emph{the
dimming of the light signal} by the gravitational wave and to
indicate that it is the high electromagnetic wave frequencies
being reflected that is responsible for the dimming.

\noindent
\section*{Acknowledgment}\noindent
C.B. and V.P.F. thank NATO for financial support under CLG/979723
and P.A.H. thanks CNRS for a position of Chercheur Associ\'e.

\appendix
\section{Collision is Head-on} \setcounter{equation}{0}
We demonstrate here that without loss of generality a frame can
always be chosen so that the collision of the plane
electromagnetic waves and the impulsive gravitational wave in
section 4 is head-on. If the plane electromagnetic waves propagate
in such a way that they collide with the gravitational wave
described by (\ref{1.1}) and (\ref{1.2}) but otherwise their
propagation direction is arbitrary then ${}^{(-)}l^\mu$ in
(\ref{3.1}) must be replaced by
\begin{equation}\label{A1}
{}^{(-)}l^{\mu}=(x_1, y_1, z_1, t_1)\ ,\qquad x_1>0\ ,\qquad
t_1=\sqrt{x_1^2+y_1^2+z_1^2}\ .\end{equation}Solving the null
geodesic equations (3.3)--(3.7) now yields for $\lambda <0$,
\begin{equation}\label{A2}
x=x_0+x_1\,\lambda\ ,\qquad y=y_0+y_1\,\lambda\ ,\qquad
z=z_0+z_1\lambda\ ,\qquad t=-x_0+t_1\,\lambda\ ,\end{equation}
where $x_0, y_0, z_0$ are constants of integration. For $\lambda
>0$ the null geodesic is given by
\begin{eqnarray}\label{A3}
x&=&x_0+(x_1 +X_1)\,\lambda +\hat X_1\ ,\\
y&=&y_0+(y_1 +Y_1)\,\lambda\ ,\\
z&=&z_0+(z_1 +Z_1)\,\lambda\
,\\
t&=&-x_0+(t_1 -X_1)\,\lambda -\hat X_1\ .\end{eqnarray} Here
\begin{equation}\label{A4}
\hat X_1=(h)_0\ ,\qquad Y_1=-C_0\,(h_y)_0\ ,\qquad
Z_1=-C_0\,(h_z)_0\ ,\end{equation} and
\begin{equation}\label{A5}
X_1=-\frac{1}{C_0}\left [y_1\,Y_1+z_1\,Z_1 +\frac{1}{2}\,\left
(Y_1^2+Z_1^2\right )\right ]\ ,\end{equation} with $C_0=x_1+t_1$
and otherwise the notation is the same as that of section 3. Thus
the tangent to the null geodesic on the future side ($\lambda >0$)
of $\lambda =0$ is
\begin{equation}\label{A6}
l^{\mu}=(x_1+X_1, y_1+Y_1, z_1+Z_1, t_1-X_1) .\end{equation}
Proceeding now as in section 3 and using $x_0, y_0, z_0, \lambda$
as coordinates, given in terms of the rectangular Cartesian
coordinates and time by (\ref{A2}) for $\lambda <0$ and by
(A.3)--(A.6) for $\lambda >0$ we find that $l^\mu$ can be written
\begin{equation}\label{A7}
l_\mu =\frac{\partial}{\partial
x^\mu}\{C_0\,x_0+\frac{1}{2}(\bar\zeta _1\,\zeta+\zeta
_1\,\bar\zeta)\}\ ,\end{equation} the line--element for $\lambda
<0$ takes the form
\begin{equation}\label{A8}
ds^2=|d\zeta
|^2+2\,d\lambda\,\left\{C_0\,dx_0+\frac{1}{2}(\bar\zeta _1d\zeta
+\zeta _1d\bar\zeta )\right\}\ ,\end{equation}and the
line--element for $\lambda >0$ can be put in the form
\begin{equation}\label{A9}
ds^2=\left |d\zeta +2\,\lambda\,\frac{\partial
Y_1}{\partial\bar\zeta}\,d\bar\zeta\right
|^2+2\,d\lambda\,\left\{C_0\,dx_0+\frac{1}{2}(\bar\zeta _1d\zeta
+\zeta _1d\bar\zeta )\right\}\ .\end{equation}Here as before
$\zeta =y_0+iz_0$, and
\begin{equation}\label{A10}
\frac{\partial Y_1}{\partial\zeta}=\frac{2\,C_0\,p}{(\zeta -\alpha
)^2}\ ,\end{equation} with $\alpha =a-ib$ and in addition $\zeta
_1=y_1+iz_1$. Now a transformation
\begin{eqnarray}\label{A11}
x_0\rightarrow x_0'&=&x_0+\frac{1}{2\,C_0}(\bar\zeta
_1\,\zeta+\zeta _1\,\bar\zeta)\ ,\\\lambda\rightarrow\lambda
'&=&\frac{1}{2}\,\lambda\,C_0\ ,\end{eqnarray} takes
({\ref{A8})--(\ref{A10}) into the form given by (\ref{3.20}),
(\ref{3.24}) and (\ref{3.25}). In this new frame of reference the
collision is therefore head--on.


\begin{thebibliography}{99}
\bibitem{Wei} S. Weinberg, \emph{Gravitation and Cosmology} (John Wiley,
New York, 1972), p. 244.
\bibitem{PR04} C. Barrab\`es and P. A. Hogan, Phys. Rev. D{\bf 70}, 107502 (2004).
\bibitem{PR01} C. Barrab\`es and P. A. Hogan, Phys. Rev. D{\bf 64}, 044022 (2001).
\bibitem{PR03} C. Barrab\`es and P. A. Hogan, Phys. Rev. D{\bf 67}, 084028
(2003).
\bibitem{AS} P. C. Aichelburg and R. U. Sexl, Gen. Rel. Grav. {\bf
12}, 303 (1971).
\bibitem{GV} J. Garriga and E. Verdaguer, Phys. Rev. D{\bf 43},
391 (1991).
\bibitem{CQG} C. Barrab\`es and P. A. Hogan, Class. Quantum Grav. {\bf 21}, 405 (2004).
\bibitem{BL} R. H. Boyer and R. W. Lindquist, J. Math. Phys. {\bf
8}, 265 (1967).
\bibitem{P} R. Penrose, ``The Geometry of Impulsive Gravitational
Waves'', in {\it General Relativity: papers in honour of J. L.
Synge}, edited by L. O'Raifeartaigh (Clarendon Press, Oxford
1972), p.101.
\bibitem{R} I. Robinson, J. Math. Phys. {\bf 2}, 290 (1961).
\bibitem{S} J. L. Synge, {\it Relativity: the general theory}
(North--Holland Publishing Company, Amsterdam 1964), p.83.
\end{thebibliography}
\end{document}